\documentclass[10pt]{iopart}

\usepackage{graphicx}
\usepackage{dcolumn}
\usepackage{bm}
\usepackage[colorlinks=true, citecolor=blue]{hyperref}
\usepackage{array}
\usepackage{float} 
\usepackage{silence}
\usepackage{cite}
\usepackage{booktabs}
\usepackage{epstopdf}
\usepackage[utf8]{inputenc}
\usepackage{ragged2e}
\usepackage{hyperref}
\usepackage[mathlines]{lineno}

\usepackage{iopams}  
\begin{document}

\title[]{Tuning spin one channel to exotic orbital two-channel Kondo effect in ferrimagnetic composites of LaNiO$_{3}$ and CoFe$_{2}$O$_{4}$}

\author{Ananya Patra$^1$$^,$$^*$, Krishna Prasad Maity$^1$, Ramesh B kamble $^1$$^,$$^2$ and V Prasad$^1$}
\address{$^1$ Department of Physics, Indian Institute of Science, Bangalore 560012, Karnataka, India}
\address{$^2$ Department of Physics, College of Engineering, Pune 411005, Maharastra, India}
\ead{$^*$ananyap@iisc.ac.in}
\vspace{10pt}
\begin{indented}
\item[]April 2018
\end{indented}

\begin{abstract}
We report the tuning from spin one channel (1CK) to orbital two-channel Kondo (2CK) effect by varying CoFe$_{2}$O$_{4}$ (CFO) content in the composites with LaNiO$_3$ (LNO) along with the presence of ferrimagnetism. Although there is no signature of resistivity upturn in case of pure LNO, all the composites exhibit a distinct upturn in the temperature range 30--80 K. For composite with lower percentage of CFO (10 \%), the electron spin plays the key role in the emergence of resistivity upturn which is affected by external magnetic field. On the other hand, when the CFO content is increased ($\geq$ 15$\% $), the upturn shows strong robustness against high magnetic field ($\leq$ 14 T) and a crossover in temperature variation from lnT to T$^{1/2}$ at the Kondo temperature, indicating the appearance of orbital 2CK effect. The orbital 2CK effect is originated due to the scattering of conduction electrons from the structural two-level systems which is created at the interfaces between the two phases (LNO and CFO) of different crystal structures as well as inside the crystal planes. A negative magnetoresistance (MR) is observed at low temperature ($<$ 30 K) for composites containing both lower (10 \%) and higher percentage (15 \%) of CFO. We have analyzed the negative MR using Khosla and Fisher semi-empirical model based on spin dependent scattering of conduction electrons from localized spins. 
\end{abstract}
%

%
%
%
\ioptwocol
\section{Introduction}
Kondo effect has been a topic of intense study due to its significant contribution to understand and develop the theory of strongly correlated systems, high T$_{c}$ superconductors, quantum dots, spintronics and heavy Fermions \cite{1,2,3,4}. In metals containing small amount of magnetic impurity, the resistivity upturn arises due to the spin-flip scattering of conduction electrons off the impurity spins. This is known as spin Kondo effect as proposed by J Kondo in 1964 \cite{5}. In this phenomenon electrons in only one channel are coupled to the impurity, so it is also referred to as spin single channel Kondo (1CK) effect. Later in 1980, Zawadowski \cite{6} and Nozi\`{e}res \cite{7} independently introduced the concept of two-channel Kondo (2CK) effect where two individual channels of electrons are equally coupled with the magnetic moment of the impurity. Potok et al has successfully provided an experimental demonstration of this 2CK model by fabricating a device of two independent reservoirs of electrons serving as the two channels and a quantum dot with odd number of electrons acting as the impurity spin \cite{8}. They have shown both spin 1CK and 2CK as the possible ground states by tuning the two different coupling constants between the impurity and the reservoirs. The coupling constants have been controlled by changing the occupancy of both finite reservoir and the quantum dot. Additionally, there are other reports on experimental manifestation of the spin 2CK effect \cite{9,10}. The 2CK effect can also be realized if the conduction electrons in the two channels are symmetrically coupled to the 
impurity with degenerate degrees of freedom other than spin, for example charge or orbital quantum number. These phenomena are called charge \cite{11,12,13} and orbital \cite{14,15} 2CK effect, respectively. 

As suggested by Zawadowski \cite{6}, the orbital 2CK effect occurs due to the interaction of conduction electrons with the scattering centres or more precisely the structural defects modelled as two-level system (TLS). This TLS can be represented by a symmetric double well potential where an atom or a group of atoms can coherently tunnel between the two potential minima at a rate of $10^{8}$ to $ 10^{12}$ Hz~\cite{16}. In this case, the orbital quantum number of the scatterers play the role of (pseudo) spin and the spin degeneracy (up and down) of conduction electrons form the two channels. The TLS may arise from the positional disorder in metallic glasses \cite{17} or due to the local distortion created as a result of Jahn-Teller effect \cite{18} or by the presence of non-magnetic disorder in the crystals \cite{19}. The orbital 2CK effect depends on the symmetry and the strength of coupling (J) between conduction electrons and the TLS, the tunnelling rate of the atom and the imbalance of electron density in the two channels. The resistivity upturn caused by orbital 2CK remains unaffected under applied magnetic field unlike the 1CK effect, since spin is not explicitly involved in the transport properties \cite{20,21}. Considering the temperature variation of resistivity, 2CK follows logarithmic behaviour just as 1CK effect upto the Kondo temperature (T$_{K}$). However below T$_{K}$, the temperature variation in 2CK model can not be explained by typical Fermi liquid behaviour ($\rho \sim T^{2}$) because the concept of quasi-particles is no longer applicable and as a result exotic non-Fermi liquid ($\rho \sim T^{1/2}$) behaviour takes place in contrast to the 1CK effect \cite{16}.

In spite of these detailed theoretical analysis, the existence of orbital 2CK effect was questionable for last few decades due to experimentally inaccessible value of $ T_{K} $ and hence absence of proper experimental evidences. $ T_{K} $ is sensitive to the coupling constant (J) and the density of states at Fermi level (N($\varepsilon _{F} $)):
$ T_{K} \propto exp(\frac{1}{JN(\varepsilon_{F})}) $. Aleiner et al \cite{22,23} have shown that 2CK can not be observed in the weak coupling regime (JN$(\varepsilon_{F}) \ll$ 1). However, the virtual electron assisted hopping of the atom to the higher excited states can lead the value of $T_{K}$ to experimentally observable range (1--10 K) as suggested by Zar\'{a}nd and Zawadowski \cite{24}. In another report Zar\'{a}nd further pointed that if the electrons interact with the TLS through resonant scattering then the strong coupling regime is reached and thus the orbital 2CK effect can be achieved \cite{21}. In addition, many groups have reported the experimental demonstration of orbital 2CK effect due to the electron scattering from structural disorder as illustrated by the TLS model \cite{18,20,25,26,27,28}. For example, Cichorek et al reported an lnT variation of resistivity in glass-like single crystal ThAsSe which is not affected by the strong magnetic field ($\leq$ 17 T) or high hydrostatic pressure ($\leq$ 1.88 GPa). The scattering of conduction electrons with the structural defects gives rise to this type of resistivity behaviour \cite{25}. They have also shown that non-magnetic 2CK effect arises due to the dynamic defects in the square nets of As in the layered compound ZrAs$_{1.58}$Se$_{0.39}$ \cite{18}. Zhu et al have reported magnetic field independent resistivity upturn and a deviation from 2CK to NFL behaviour around T$_{K}$ in ferromagnetic thin films  L1$_{0}$-MnGa and L1$_{0}$-MnAl \cite{14,15}. An antiparallel alignment between Mn-Mn atoms due to the superexchange coupling is the reason of coexistence of 2CK fixed point with ferromagnetism.

In order to get new insights and to deepen our knowledge about the coexistence of magnetism and the orbital 2CK effect, we have extensively studied the low temperature magneto-transport properties of the composites containing LaNiO$_{3}$ (LNO) and CoFe$_{2}$O$_{4}$ (CFO) [(1-x)LNO + xCFO; x = 0, 0.10, 0.15, 0.20, 0.25)]. LNO is an exceptional metal oxide with high electrical conductivity $\sim$ 10$^{5} \Omega ^{-1}$-m$^{-1}$ and it is paramagnetic in nature \cite{29,30}. CFO is well known as hard ferrimagnetic material with high coercivity ($\sim$ 750-980 Oe), high Curie temperature ($\sim$ 520$^{o}$C) and good chemical stability \cite{31,32}. We have observed that a pronounced resistivity upturn at low temperature is the salient feature for all the composites. However, it does not appear for pure LNO. According to earlier reports, the resistivity upturn for LNO thin film is caused due to the weak localization \cite{33} whereas for polycrystalline film \cite{34} or nanostructured LNO \cite{35}, the Anderson localization dominates at the grain boundaries. Therefore the absence of upturn in pure LNO indicates the negligible effect of disorder induced localization on the electron conduction. We have shown that the low temperature transport property of composite containing lower percentage of CFO (10\%) is dominated by the spin 1CK effect. On the other hand, with increasing CFO content ($\geq$ 15\%) the conduction mechanism is governed by the orbital 2CK effect as confirmed by these two important observations: A crossover from lnT to T$^{1/2}$ variation around T$_{K}$ signifying non-Fermi liquid (NFL) behaviour, appeared only for overscreening impurity in case of 2CK effect. In addition, the temperature variations of resistivity as lnT and T$^{1/2}$ are essentially independent of the high magnetic field ($\leq$ 14 T).

Due to the high value of T$_{K}$ achieved in our composites, it is easy to reach the orbital 2CK effect as well as the NFL behaviour experimentally and hence it helps to provide a deeper insight in the physics of 2CK fixed point. By varying impurity (CFO) content we can give rise to a crossover from spin 1CK to orbital 2CK effect along with the presence of ferrimagnetism. Therefore the present study of the composites can help us to understand the effect of spin as well as structural defects created by the magnetic impurity on the transport and magnetic properties.

\section{\label{sec:level2}Experimental Details}
The polycrystalline composites of LNO and CFO [(1-x)LNO + xCFO (x= 0.10, 0.15, 0.20, 0.25)] are prepared via standard solid state reaction method. LNO and CFO nanoparticles (nps) are synthesized individually via citric acid assisted sol-gel method \cite{35,36}. Then the two powders are mixed and ground thoroughly in required ratio and heated at 650 $ ^{0} $C in air. For low temperature transport property measurements, the heat treated powder is pressed into pellet and sintered at 700 $ ^{0} $C in air. 
All the samples are characterized by X-ray diffraction (XRD), Scanning electron microscopy (SEM) and Tunneling electron microscopy (TEM) techniques. XRD has been performed using Rigaku Smartlab diffractometer. SEM and TEM images are taken with Sirion XL30 FEG SEM and FEI Tecnai F30 S-TWIN TEM respectively. Magnetic and magneto-transport properties have been measured in Quantum Design PPMS 9 T and 14 T respectively down to the lowest reachable temperature 10 K.

\section{\label{sec:level3}Results and Discussion}

\subsection{\label{sec:level4}X-ray Diffraction}

\begin{figure}[htb]
	\centering
	\includegraphics[width=1\columnwidth]{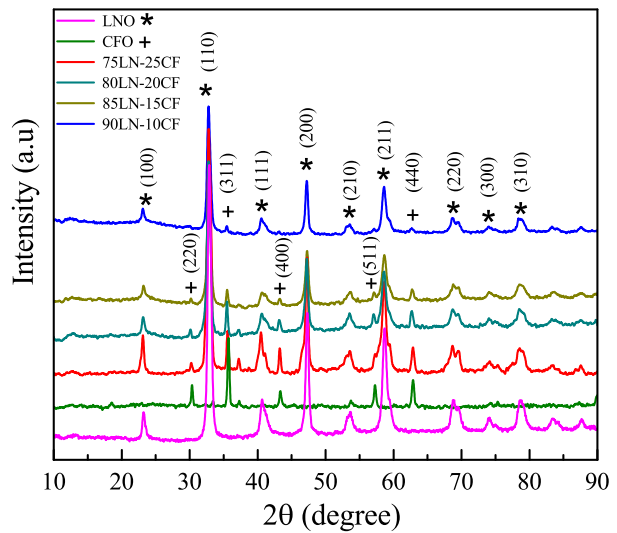}
	\small{\caption{XRD pattern of the composite materials for the series of (1-x)LNO + xCFO (x= 0, 0.10, 0.15, 0.20, 0.25 and 1)\label{fig.1}}}
\end{figure} 
We have confirmed the phase formation of pure LNO and CFO and the phase purity of the composites for the series of (1-x)LNO + xCFO (x= 0.10, 0.15, 0.20, 0.25) by XRD pattern, as shown in~\Fref{fig.1}. Both LNO and CFO peaks co-exist independently in the composites as indicated by * and + respectively. The diffraction peaks of LNO are indexed to the Rhombohedral perovskite structure (JCPDS No.~33-0711) and that of CFO are related to cubic spinel structure (JCPDS No.~22-1086). The XRD peaks of CFO corresponding to the planes (220), (400) and (511) appear in the composites only when the CFO percentage is $\geq$ 15. We could not witness any extra peaks which confirm that there is no chemical reaction between the elements and the absence of impurity phases. The sharp peaks of LNO and CFO in the composites imply their intact crystallinity.

\subsection{\label{sec:level5}SEM and TEM}
\begin{figure}[htb]
	\centering
	\includegraphics[width=1\columnwidth]{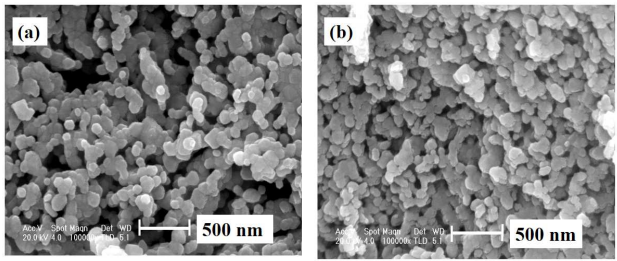}
	\small{\caption{SEM images for the composites with (a) x= 0.10 and (b) x= 0.20\label{fig.2}}}
\end{figure}

The morphology of the composites have been studied using scanning electron microscopy (SEM) for the two different values of x (0.10 and 0.20) as shown in~\Fref{fig.2} which reveal the coarse-grained structure of the composites. There is not much variation in the average grain size for the different CFO content as evident from~\Fref{fig.2} (a) and (b). The grain sizes are 78 nm and 76 nm for x = 0.10 and 0.20 respectively.

\begin{figure}[htb]
	\centering
	\includegraphics[width=0.8\columnwidth]{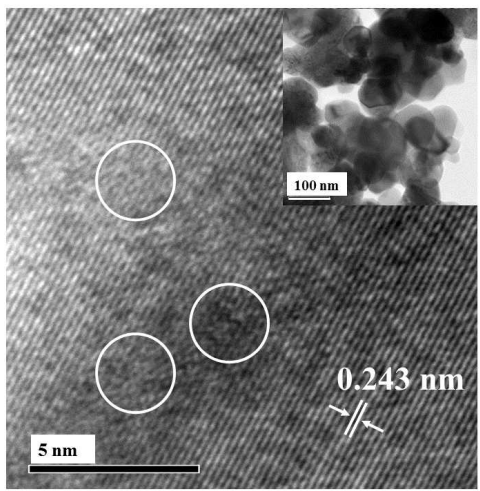}
	\small{\caption{HR-TEM images of 80LN-20CF. The white circles point out the dislocation portion in (311) plane of CFO. Inset shows the nanocrysttaline composite at low resolution. \label{fig.3}}}
\end{figure}

\Fref{fig.3} displays the High resolution TEM (HR-TEM) image for x = 0.20 with inset showing TEM image of the nanocrystalline composites. In HR-TEM image, the lattice spacing is 0.243 nm which corresponds to the (311) plane of CFO (JCPDS No.~22-1086). The highlighted white circles indicate the presence of dislocations which are possibly generated from the local strain created due to the lattice mismatch or the different thermal expansion coefficients of LNO ($\sim$ 10 $\times$ 10$^{-6}$ K$^{-1}$) \cite{37} and CFO ($\sim$ 14.9 $\times$ 10$^{-6}$ K$^{-1}$) \cite{38}. 

\subsection{\label{sec:level6}Magnetic properties}

\begin{figure}[htb]
	\centering
	\includegraphics[width=1\columnwidth]{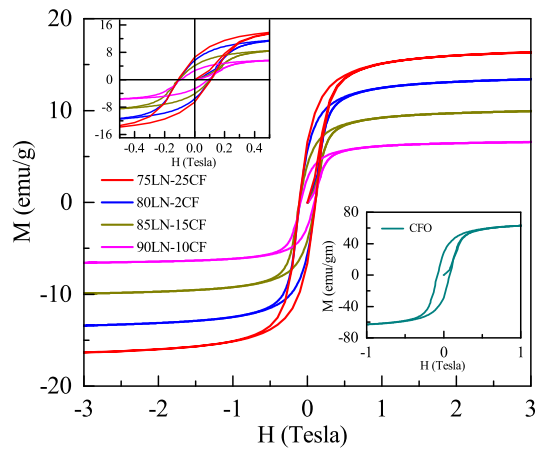}
	\caption{Magnetic hysteresis curves measured at room temperature for the composites with x = 0.10, 0.15, 0.20, 0.25 with the top inset showing the enlarge view at low magnetic field and the bottom inset showing the M-H loop for pure CFO\label{fig.4}}
\end{figure}

To probe the magnetic properties we have measured the hysteresis loops (M vs. H) at room temperature for the nano-composite materials (\Fref{fig.4}) and for pure CFO (bottom inset of~\Fref{fig.4}) which reflect the ferrimagnetic behaviour of all samples. The values of coercive field (H$ _{c}$), saturation magnetization (M$ _{s}$) and remanence magnetization (M$ _{r}$) are summarized in~\Tref{table 1} which shows all the parameters decrease consistently with decrease in CFO content. The magnetic parameters of pure CFO are close to the earlier reported values \cite{31,32}. The ferrimagnetic behaviour of pure CFO as well as of the composites are intact down to the lowest temperature 10 K as we have confirmed from the magnetization vs. temperature curve in the field cooled and zero-field cooled measurement \cite{39} (Supplementary material).

\begin{table}[htbp]
	\setlength{\tabcolsep}{6pt} 
	\renewcommand{\arraystretch}{1.2} 
	\centering
	\small{\caption{The Magnetic parameters of pure CFO and the composites at room temperature\label{table 1}}}
	\resizebox{\linewidth}{!}{%
		\begin{tabular}{lccc}
			\toprule[0.5pt]
			Sample &  M$_{s}$ (emu/g)&  M$_{r}$ (emu/g) & H$_{c}$ (Tesla) \\
			\midrule[0.2pt]
			CFO       & 67.6 &  28.7 & 0.107 \\
			75LN-25CF & 16.5 &  6.5 & 0.099 \\
			80LN-20CF & 13.6 &  5.5 & 0.098 \\
			85LN-15CF & 10.1 &  4.0 & 0.085 \\
			90LN-10CF &  6.7 &  2.5 & 0.078 \\
			\bottomrule[0.5pt]
		\end{tabular}}
	\end{table}

\subsection{\label{sec:level7}Temperature dependence of resistivity}
\begin{figure}[htb]
	\centering
	\includegraphics[width=1\columnwidth]{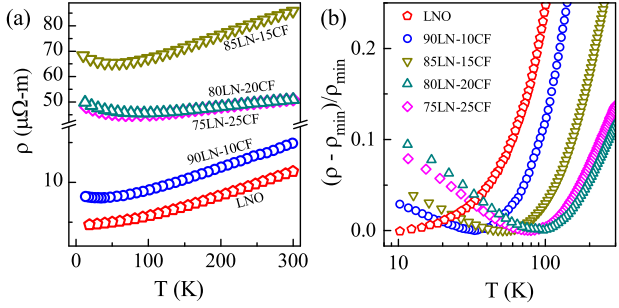}
	\small{\caption{At zero magnetic field, (a) resistivity and (b) semi-log plot of normalized resistivity ($(\rho-\rho_{min})/\rho_{min}$) as a function of temperature for pure LNO and the composites (x = 0.10, 0.15, 0.20 and 0.25).\label{fig.5}}}
\end{figure}

\begin{table*}[htbp]
	\setlength{\tabcolsep}{10pt} 
	\renewcommand{\arraystretch}{1.4} 
	\centering
	\caption{Comparison of $T_{min}$ and depth of minimum ($\delta\rho= (\rho _{10K} $ - $ \rho_{min} $)/$ \rho_{10K} $) for the composites with x = 0.10, 0.15, 0.20 and 0.25 at various magnetic fields \label{table 2}}
	\resizebox{0.9\textwidth}{!}{%
		\begin{tabular}{*9c}
			\toprule[0.5pt]
			
			Sample & \multicolumn{4}{c}{$T_{min}$  (K)} & \multicolumn{4}{c}{Depth of minimum ($\delta\rho$)}\\
			\midrule[0.2pt]
			
			{}  & 0 T   & 4 T    & 8 T   & 14 T   & 0 T   & 4 T    & 8 T   & 14 T \\
			\cmidrule[0.2pt](l{1em}r{1em}){2-5}  \cmidrule[0.2pt](l{1em}r{1em}){6-9}
			
			90LN-10CF   &  33.10 & 34.21   & 33.08  & 33.36 &  0.020 & 0.019   & 0.016  & 0.012\\
			
			85LN-15CF   &  50.20 & 51.64 & 51.81   & 53.18 &  0.043  &  0.041   & 0.039  & 0.036 \\

			80LN-20CF   &  89.08 & 87.06 & 82.45   & 90.47 &  0.086  &  0.087   & 0.084  & 0.080 \\
			
			75LN-25CF   &  79.68  &  80.07  & 75.94  & 76.32 &  0.077  &  0.076   & 0.074  & 0.069\\
			
			\bottomrule[0.5pt]
		\end{tabular}}
		
	\end{table*}
We have measured the resistivity ($\rho$) as a function of temperature (T) from 300 K to 10 K for the composites with different content of CFO (x = 0, 0.10, 0.15, 0.20 and 0.25) as shown in~\Fref{fig.5} (a). It reveals that the resistivity of pure LNO decreases continuously with decreasing temperature without exhibiting any signature of resistivity upturn, whereas a distinct resistivity minimum ($\rho_{min}$) is observed for all the composites in the temperature range from 30 K to 80 K at zero magnetic field. To clearly visualize the upturn we have plotted the normalized resistivity ($(\rho-\rho_{min})/\rho_{min}$) in~\Fref{fig.5} (b). The temperature (T$ _{min} $) at which resistivity minimum occurs and the depth of minimum, defined as $\delta\rho= (\rho _{10K} $ - $ \rho_{min} $)/$ \rho_{10K} $ strongly depend on the CFO content and get enhanced with increasing CFO percentage as observed from~\Tref{table 2}. In the following sections, we will discuss about the transport properties of pure LNO and then composites with x = 0.10 (90LN-10CF) and x $\geq$ 0.15 in details. For all the samples, resistivity is recorded as a function of temperature from  300 K to 10 K at different magnetic fields (0 T, 4 T and 8 T). Moreover, to observe the effect of high magnetic field on $\rho_{min}$, low temperature data is taken at 14 T, the highest accessible field in our experiment. In all measurements, the magnetic field is applied perpendicular to the flow of current.

\subsubsection{\label{sec:level8}Pure LNO:}
\begin{figure}[htb]
	\centering
	\includegraphics[width=1\columnwidth]{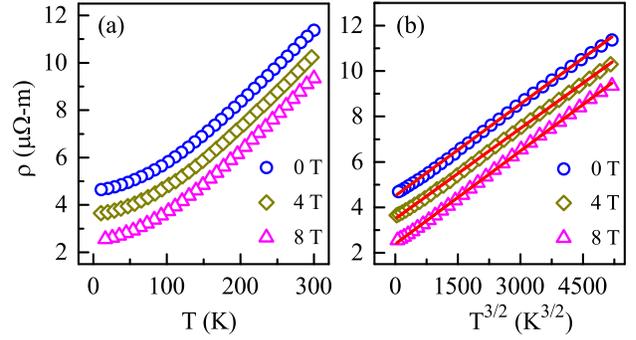}
	\small{\caption{(a) $\rho$ vs.~T and  
			(b) $\rho$ vs.~T$^{3/2}$ for pure LNO  at different external magnetic fields. (For a clear view, the curves at non zero fields are vertically shifted by few $ \mu \Omega $-m.)\label{fig.6}}}
\end{figure}
The temperature dependence of resistivity of pure LNO is shown in Fig.~\ref{fig.6} (a). The resistivity follows a power law with T$^{3/2}$ in the whole temperature range (300--10 K) with and without applying magnetic field (\Fref{fig.6} (b)). Similar type of temperature variation has been reported earlier in bulk LNO polycrystal \cite{40,41} as well as in LNO superlattices or heterostructures (eg. LaNiO$ _{3} $/SrTiO$ _{3}$ \cite{42,43,44}, LaNiO$ _{3} $/SrMnO$ _{3}$ \cite{45} and LaNiO$_{3}$/LaAlO$ _{3}$  \cite{40}) where T$^{3/2}$ dependence is emanated from the localized antiferromagnetic spin fluctuations \cite{41,42,45}. In bulk LNO, these localized spins might be originating from Ni$ ^{+2} $ ions created due to the Oxygen deficiency \cite{41} whereas for superlattice structures, this spin fluctuation arises because of the formation of spin density wave in the heterostructure interfaces~\cite{42,46}. Another possible reason behind this T$^{3/2}$ variation could be the combination of electron-electron ($\sim$ T$^{2}$) and electron-phonon ($\sim$ T) interaction~\cite{47,48,49}. 

In present case, we observe that the slope of the straight line ($\rho$ vs. T$^{3/2}$) does not vary with external magnetic field (\Fref{fig.6} (b)). Since both electron-electron and electron-phonon interactions are magnetic field independent phenomena, therefore T$^{3/2}$ variation can be ascribed to the combined effect of these two interactions. Next we have discussed the effect of adding ferrimagnetic CFO on the transport property of conducting LNO.

\subsubsection{\label{sec:level9}90LN-10CF:}

\begin{figure*}[htb]
	\centering
	\includegraphics[width=1\textwidth]{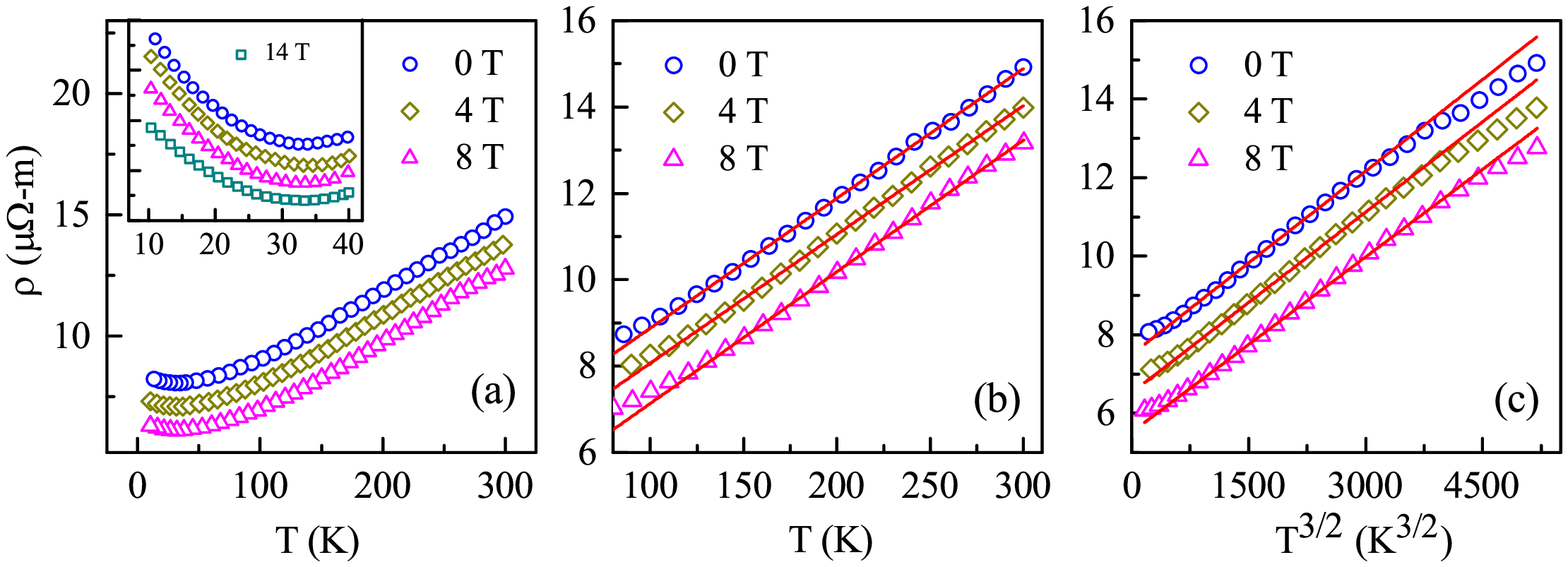}
	\includegraphics[width=1\textwidth]{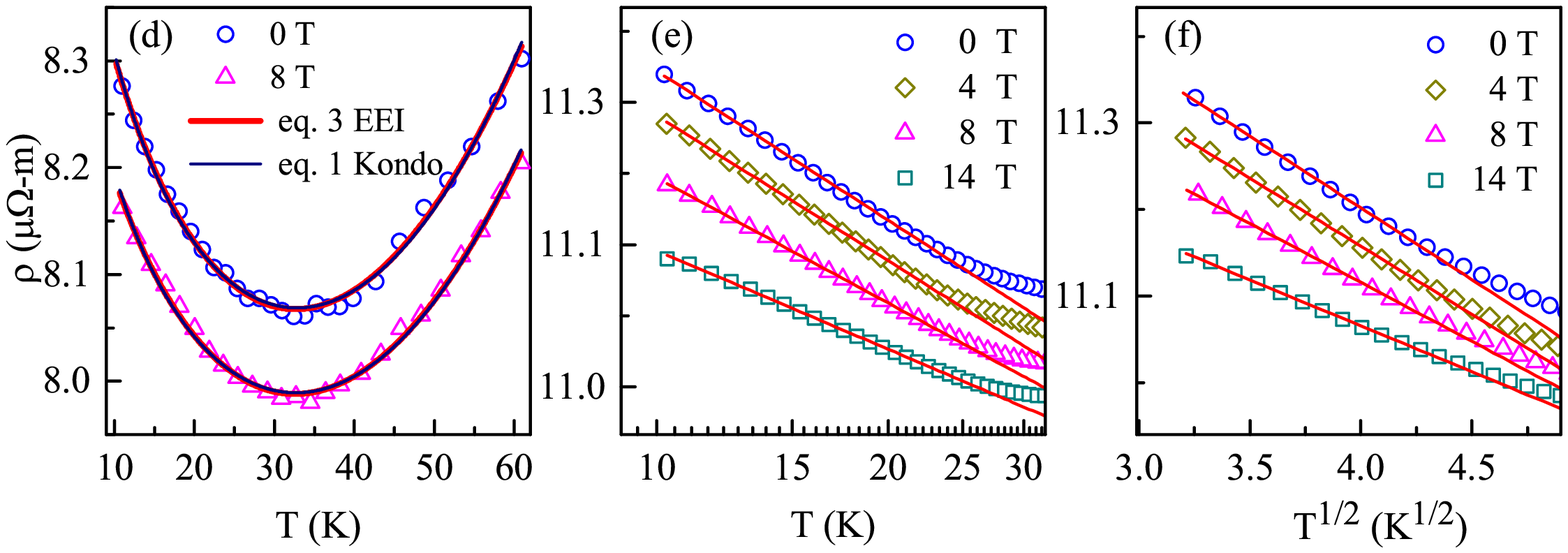}
	\caption{For composite with x = 0.10, (a) $\rho$ vs.~T at different magnetic fields. Inset shows magnified low temperature region to clearly visualize the resistivity upturn. $\rho$ fitted at different temperature regions: (b) $\rho$ vs.~T (300--140 K), (c) $ \rho $ vs.~T$ ^{3/2}$ (75--215 K), (d) the resistivity upturn fitted with Eqs.~\ref{eq.1} and \ref{eq.3} at 0 T and 8 T, (e) semilog plot of $\rho$ vs.~T (Kondo) (10--24 K) and (f) $ \rho $ as a function of T$ ^{1/2}$ (EEI) (10--19 K).(For a clear view, the curves at non zero fields are vertically shifted by few $ \mu \Omega $-m.).\label{fig.7}}
\end{figure*}

The resistivity as a function of temperature for composite with x = 0.10 is presented in~\Fref{fig.7} (a). \Fref{fig.7} (b) shows a linear temperature dependence of resistivity from 300 K to 140 K due to collisions between thermally excited phonons and itinerant electrons. With decreasing temperature, the linear T dependence deviates and T$^{3/2}$ variation occurs down to 75 K, similar type of behaviour as pure LNO (\Fref{fig.7} (c)).

On further decreasing temperature, the resistivity exhibits a minimum at $\sim$ 33 K. In order to comprehend the origin of resistivity minimum, the experimental data have been analyzed considering three main mechanisms which are accountable for resistivity upturn: (a) Spin polarized tunneling through grain boundaries, (b) Kondo effect based on spin dependent scattering and (c) Quantum interference effect (QIE). Weak localization and electron-electron interaction (EEI) are the two leading contributions of QIE to resistivity correction at low temperature.

Usually in polycrystalline sample, the grain boundary tunneling plays a key role in the low temperature transport properties. With lowering temperature, the movement of spins through grain boundaries is inhibited due to the spin freezing and thereby spins are confined in the individual grains. Consequently the resistivity starts rising below a certain temperature. But on applying magnetic field, the spins are aligned towards the field direction which assists to reduce the confinement of the spin polarized charge carriers and to increase the tunneling probability of the carriers through grain boundaries. Hence, magnetic field weakens the resistivity minimum and gives rise to a flat resistivity after a certain critical field \cite{50,51}. Inset of~\Fref{fig.7} (a) shows the resistivity as a function of temperature in zero field and three external magnetic fields of 4 T, 8 T and 14 T and the corresponding values of  $T_{min}$ and $\delta\rho$ are listed in~\Tref{table 2}. It is clear that even very high magnetic field (14 T) can not suppress the resistivity minimum. Therefore, we can rule out the possibility of grain boundary tunneling.

The weak localization arises from the quantum interference effect between two electronic waves. The magnetic field reduces the interference effect due to the destruction of wave coherence and hence suppresses the resistivity minimum \cite{52}. Therefore, the weak localization effect can as well be ruled out. Another reason to exclude the weak localization effect in our samples is the dimensionality, since in case of bulk samples or thick films, weak localization term is less important \cite{53}.

Next we will analyze $\rho_{min}$ considering other two possible mechanisms, the one channel Kondo effect (1CK) and EEI. The spin 1CK effect gives rise to a logarithmic variation of resistivity with temperature considering second order perturbation term \cite{5,54}. So the total resistivity can be written as:
\begin{equation}
\centering
\rho = \rho_{0}+\rho_{m}T^{3/2}+\rho_{p}T^{3}-\rho_{k}ln(T)
\label{eq.1} 
\end{equation}

where $ \rho_{0} $ is the residual resistivity originating from the lattice defects and impurities \cite{55},  $ \rho_{m} $ is the coefficient of T$^{3/2}$ variation as described previously, $ \rho_{p} $ denotes the phonon assisted s-d electron scattering at low temperature in the transition metals \cite{56,57} and  $\rho_{k}$ represents the contribution from Kondo effect.

The EEI plays dominant role in the strongly correlated systems and is enhanced by the strong disorder potential \cite{53}. Altshuler and Aronov first pointed out that EEI accompanied with the impurity scattering leads to a singularity in the density of states near Fermi level that causes an anomalous temperature dependence of resistivity \cite{58}. According to localization theory in a disordered system, EEI term varies as $-T^{1/2}$ in 3D regime. Considering Altshuler and Aronov calculations and including Hartree terms, its coefficient is given by the following expression \cite{59,60}:
\begin{equation}
\rho_{e}= \rho_{0}^{2}\hspace{1mm}\frac{e^{2}}{4\pi^{2}\hbar}\hspace{1mm}\frac{1.3}{\sqrt{2}}\hspace{1mm}\Big(\frac{4}{3}-\frac{3}{2}\tilde{F}_{\sigma}\Big)\hspace{1mm}{\Big(\frac{k_{B}}{\hbar D}\Big)^{1/2}}
\label{eq.2} 
\end{equation}

Here,the parameter $\tilde{F}_{\sigma}$ is the screening constant for Coulomb interaction and D is the diffusion constant.

So the total resistivity takes the form:
\begin{equation}
\rho = \rho_{0}+\rho_{m}T^{3/2}+\rho_{p}T^{3}-\rho_{e}T^{1/2}
\label{eq.3}
\end{equation}

In~\Eref{eq.1} and \Eref{eq.3}, all the coefficients are positive as sign convention has already been taken into account. In \Fref{fig.8} (d), the data at zero field and external magnetic field (8 T) are well fitted using both equations. For further insight, we have plotted the data in the region 10--20 K with lnT (\Fref{fig.7} (e)) and T$^{1/2}$ (\Fref{fig.7} (f)) which exhibit straight lines. Following the work by Xu et al \cite{50}, we attempted to fit the data considering the terms of Kondo effect and EEI simultaneously. But it results in positive contribution from either from these two effects which is not correct from theoretical point of view. Hence only from curve fitting we can not conclude which one is responsible for resistivity upturn. To distinguish between these two transport mechanisms, we have measured magnetoresistance (defined as $\bigtriangleup\rho/\rho= (\rho(H)-\rho(0))/\rho(0)$) up to 8 T at low temperature. Since Kondo effect depends on the spin dependent scattering, so in the presence of high magnetic field, due to the alignment of impurity spins along the direction of the field, spin flip scattering is suppressed which gives rise to negative magnetoresistance (MR) \cite{61}. On the other hand, EEI causes positive MR due to the splitting of spin-up and spin-down bands and the orbital effects \cite{59,62,63}.

The low temperature MR data shown in~\Fref{fig.11} (a) is found to be negative which in turn confirms the Kondo effect. The negative MR has been fitted using Khosla Fisher model as explained in~\Sref{sec:level13}.

\subsubsection{\label{sec:level10}(1-x)LN-xCF (x = 0.15, 0.20, 0.25): Two channel Kondo effect}

\begin{figure*}[htb]
	\centering
	\includegraphics[width=1\textwidth]{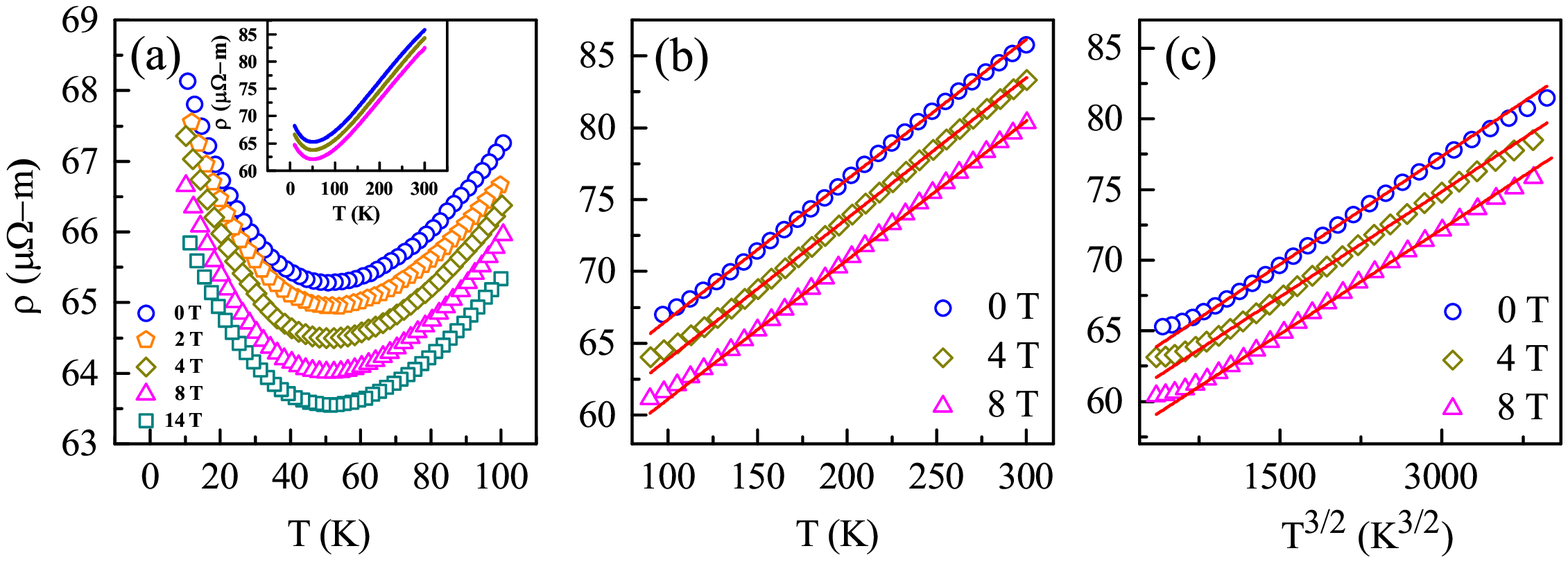}
	\includegraphics[width=1\textwidth]{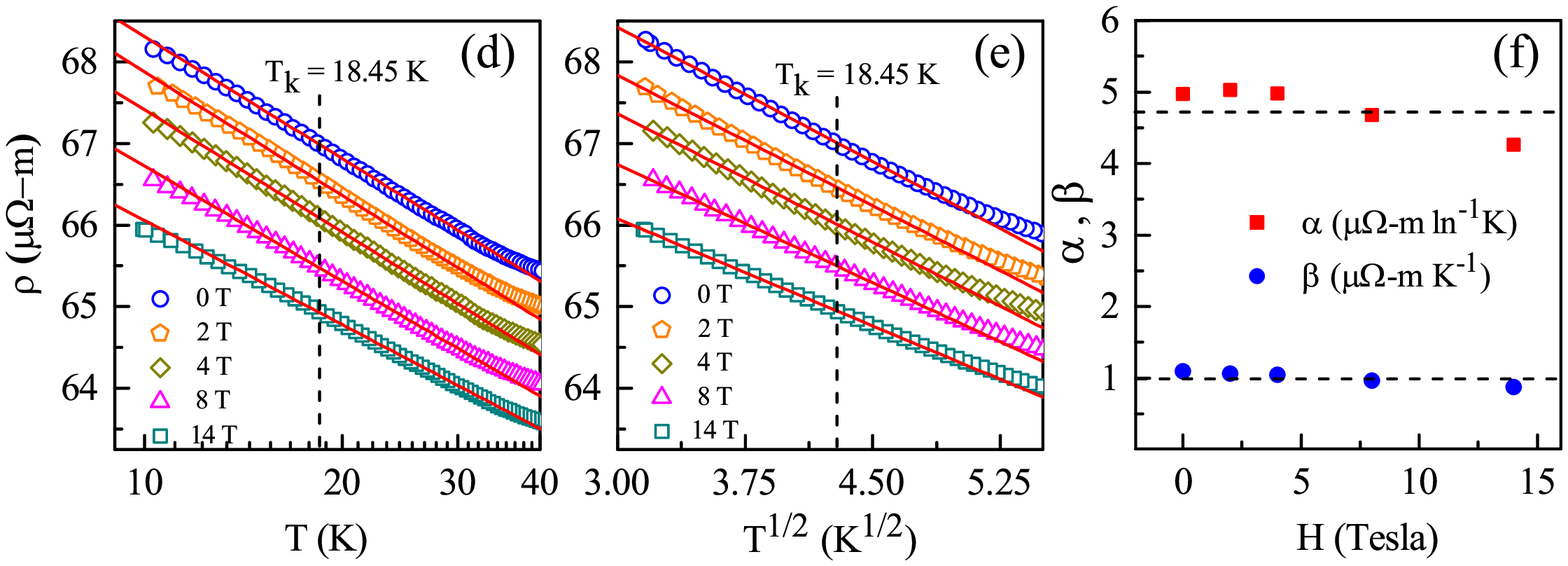}
	\small{\caption{For composite with x = 0.15, (a) the resistivity upturn at low temperature at various magnetic fields (0 T, 2 T, 4 T, 8 T and 14 T). Inset displays the nature of $\rho$ in the whole temperature range (300--10 K). $\rho$ fitted at four different temperature regions: (b) $\rho$ vs.~T (300--125 K), (c) $\rho$ vs.~T$^{3/2}$ (210--90 K), (d) semi-log plot of $\rho$ vs.~T (30--13 K), (e) $\rho$ vs.~T$^{1/2}$ (20--10 K). (For a clear view, the curves at non zero fields are vertically shifted by few $ \mu \Omega $-m) and (f) plot of $\alpha$ and $\beta$ (defined in the main text) as a function of magnetic field. \label{fig.8}}}
\end{figure*}

When we increase the percentage of CFO in the composite, the low temperature resistivity upturn becomes more prominent as the values of $T_{min}$ and $\delta\rho$ get enhanced with CFO (Table~\ref{table 2}). Here, we have discussed the transport properties of only one composite 85LN-15CF in details because the other composites with higher value of x ($>$ 0.15) follow similar type of behaviour. \Fref{fig.8} (a) displays the resistivity minimum of 85LN-15CF occurring at $\sim$ 50 K with inset showing the temperature variation of resistivity from 300 K to 10 K at different magnetic fields.

At high temperature, the resistivity is influenced by the electron-phonon and the electron-electron scattering similar to the composite with x = 0.10, as also evident from the ~\Fref{fig.8} (b) and (c). However at low temperature  (T $<$ T$_{min}$), it is intriguing to notice that the electron conduction follows remarkably different mechanism than the spin 1CK and the origin of resistivity upturn is ascribed to orbital two-channel Kondo (2CK) effect. Below $T_{min}$, $ \rho $ first varies linearly with lnT from a temperature denoted by $ T_{o}$. On further decreasing the temperature, $\rho $ deviates from lnT and a crossover to T$^{1/2}$ variation occurs at a certain temperature, referred to as the Kondo temperature (T$_{K} $). With reference to the 1CK effect  where the impurity is fully screened by the conduction electrons, the logarithmic temperature dependence should saturate following a temperature variation of T$^{2}$, as expected from the Fermi liquid theory. On the contrary, $T^{1/2}$ dependence exclusively signifies the exotic non-Fermi liquid behaviour, appeared when the impurity is overcompensated by the excess of conduction electrons and in the marginal case of this overcompensated process, the 2CK effect occurs \cite{16}. The emergence of orbital 2CK is also confirmed by the magnetic field independent nature of resistivity. We have measured $\rho$ as a function of temperature in the presence of four external magnetic fields (2 T, 4 T, 8 T and 14 T), shown in~\Fref{fig.8} (a). The experimental data is well fitted with lnT (\Fref{fig.8} (d)) and $T^{1/2} $ (\Fref{fig.8} (e)) in the temperature range  30 ($ T_{o} $)--13 K and 20--10 K respectively, for both with and without applying magnetic field. The slopes of these two straight lines are denoted by $\alpha=-d\rho/d(lnT)$ and $\beta=-d\rho/d(T^{1/2}$), respectively. There is no observable change in the values of $ \alpha $ and $ \beta $ even after applying high magnetic field of 14 T, as plotted in~\Fref{fig.8} (f). At this point we must note that this behaviour is unlike to that of the composite with x = 0.10, where the slopes start to decrease at high magnetic field ($\geq$ 8 T) (\Fref {fig.8} (e) and (f)). We have determined the value of $ T_{K} $ to be 18.45 K by considering the midpoint of the overlapping regions of the two temperature variations, lnT and $ T^{1/2} $, following the work by Zhu et al \cite{14}. The values of $ T_{o} $ and $ T_{K} $ also remain unaltered in the presence of magnetic field. This magnetic field independent scenario can arise in the two frameworks, one is orbital 2CK effect and the other one is EEI with the screening constant $\tilde{F} _{\sigma}$ (mentioned in~\Eref{eq.2}) tends to zero \cite{18}. In the following we will give explanation that strongly supports the existence of orbital 2CK in our system.

In case of EEI, the transition of temperature variation from lnT to $ T^{1/2} $ is possible if there is a dimensional crossover from 2D to 3D \cite{59}. Considering our present system which is 3D polycrystalline sample with thickness $\sim$ 1 mm, the thermal length or the inelastic scattering length is unlikely to reach to this high value. Also the high crossover temperature (T$ _{K} \sim $ 18.45 K) makes it more improbable. Hence we can easily figure out that such dimensional crossover is not feasible in our system. In addition, magnetic field has pronounced effect on EEI excluding the case when the screening constant ($\tilde{F}_{\sigma} $) is nearly equal to zero. Following the calculation of Altshuler and Aronov, we have estimated the value of $\tilde{F}_{\sigma} $ to be 0.930 which indicates a strong screening interaction (Supplementary material) \cite{39}. Therefore $\tilde{F}_{\sigma} \sim$ 0 is also not possible in our systems. Hence, we can convincingly claim that orbital 2CK effect governs the low temperature transport properties in the composites with x $\geq$ 0.15. 

The values of T$_{K}$, $\alpha$ and $\beta$ are summarized in~\Tref{table 3}. T$_{K}$ increases with the amount of CFO signifying an increasing strength of the coupling between conduction electrons and the TLS. Also the present composite materials show significantly higher value of T$_{K}$ in the range of 18--24 K with concomitant NFL behaviour.

\begin{table*}[htb]
	\setlength{\tabcolsep}{12pt} 
	\renewcommand{\arraystretch}{1.4} 
	\centering
	\caption{Comparison of $T_{K}$, $\alpha$ and $\beta$ for composites with varying CFO content (x = 0.15, 0.20 and 0.25) without and with applying constant magnetic field (4 T, 8 T and 14 T). (T$_{K}$ value does not change with magnetic field).\label{table 3}}
	\resizebox{0.9\textwidth}{!}{%
		\begin{tabular}{cccccccccc}
			\toprule[0.5pt]
			Sample & T$_{K}$ (K) & \multicolumn{4}{c}{$\alpha (\mu\Omega-m/lnK)$} & \multicolumn{4}{c}{$\beta (\mu\Omega-m/K^{1/2})$}\\
			\midrule[0.2pt]
			{}  & {} & 0 T & 4 T & 8 T & 14 T  & 0 T   & 4 T & 8 T & 14 T \\
			\cmidrule[0.2pt](l{1em}r{1em}){3-6}  \cmidrule[0.2pt](l{1em}r{1em}){7-10}
			85LN-15CF   &  18.45 & 4.97   & 4.97  & 4.67 &  4.26 & 1.09   & 1.04   & 0.96 & 0.87 \\
			80LN-20CF   &  24.32 & 5.99 & 5.89   & 5.80 &  5.47  &  1.08   & 1.06  & 1.02  & 0.94 \\
			75LN-25CF   &  22.50  &  5.32  & 5.19  & 5.04 &  4.75  &  0.98   & 0.94  & 0.91  & 0.84 \\
			
			\bottomrule[0.5pt]
		\end{tabular}}
	\end{table*}

To get an idea about the concentration of TLS ({$N_{TLS}$) in the composite 85LN-15CF, we have calculated $N_{TLS}$ using the formula \cite{16}:
	\begin{equation}
	N_{TLS} \sim \frac{\Delta\rho_{m}}{\rho} \frac{N(\varepsilon_{F})}{\tau_{e}}
	\label{eq.4}
	\end{equation}
	
where $\Delta\rho_{m}  \approx \rho(10K)- \rho_{min}$ is the maximum resistivity upturn at 0 T, $N(\varepsilon_{F})$ is the density of states at the Fermi level ($\varepsilon_{F}$), $\tau_{e}$ is calculated from the expression: $\tau_{e} = \frac{m^{*} \rho }{n e^{2}}$
where m$^{*}$ is the effective mass, n is the number density of electrons and $\rho$ is taken as the resistivity at 300 K.
Taking $\frac{\Delta\rho_{m}}{\rho} \approx$ 0.045 at 0 T for 85LN-15CF, using the values of $N(\varepsilon_{F})$, n and m$^{*}$ same as pure LNO, 1.1 $\times$ 10$^{23}$ eV$^{-1}$ cm$^{-3}$, 1.7$\times$ 10$^{23}$ cm$^{-1}$ and  11m$_{e}$ respectively \cite{64} and $\rho \approx$ 8.5 $\times$ 10$^{-3}$ $\Omega$-cm, we obtained N$_{TLS} \sim 1.67\times 10^ {23}$ cm$^{-3}$. 

\subsection{\label{sec:level11}Possible origin of TLS}

In our system there is high possibility of TLS formation due to the different crystal structures of the constituent materials (LNO and CFO). From XRD data, we have seen that LNO is a perovskite with rhombohedral structure whereas CFO belongs to the cubic spinel group.

\begin{figure}[htb]
	\centering
	\includegraphics[width=0.9\columnwidth]{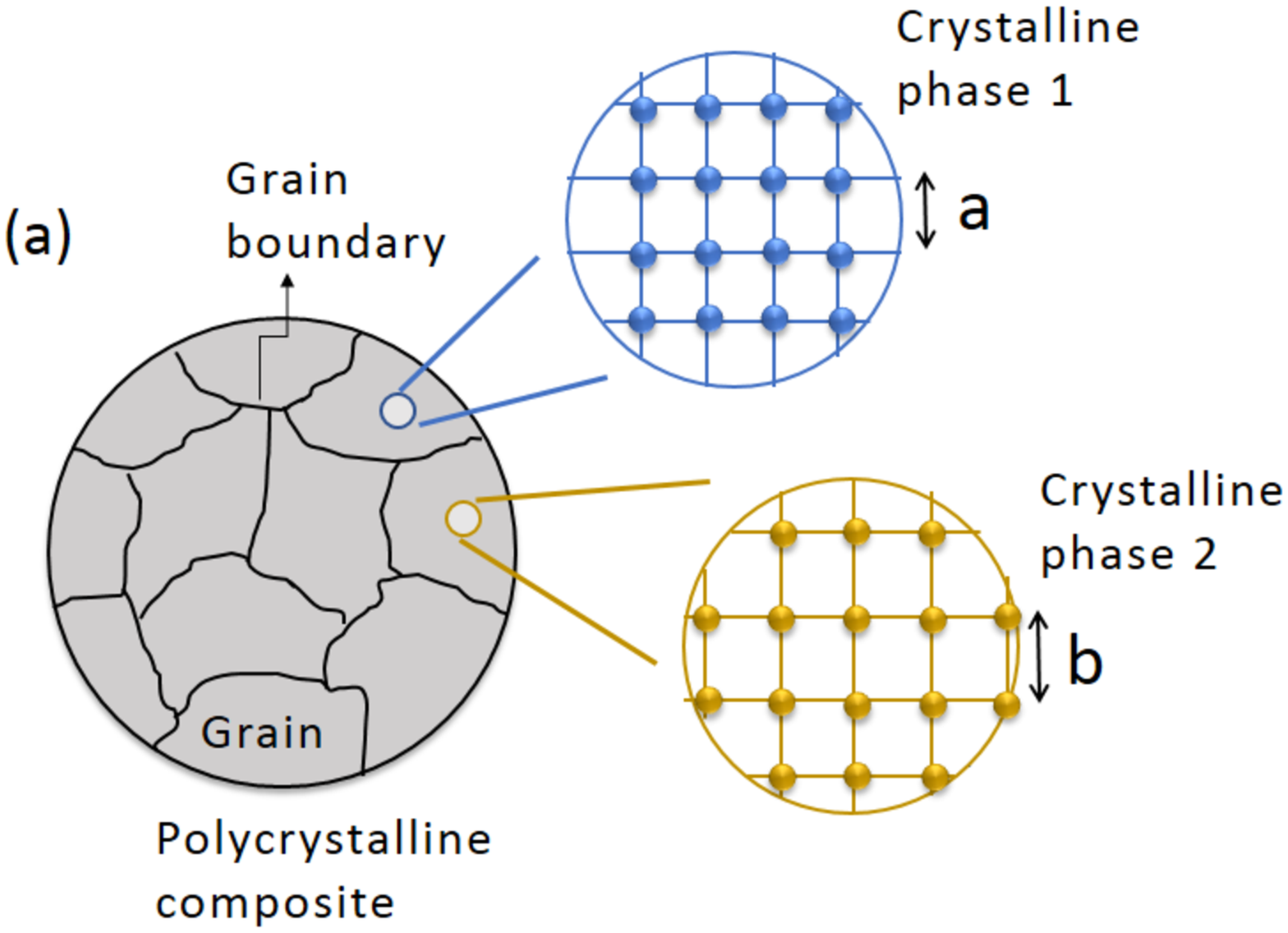}
	\includegraphics[width=1\columnwidth]{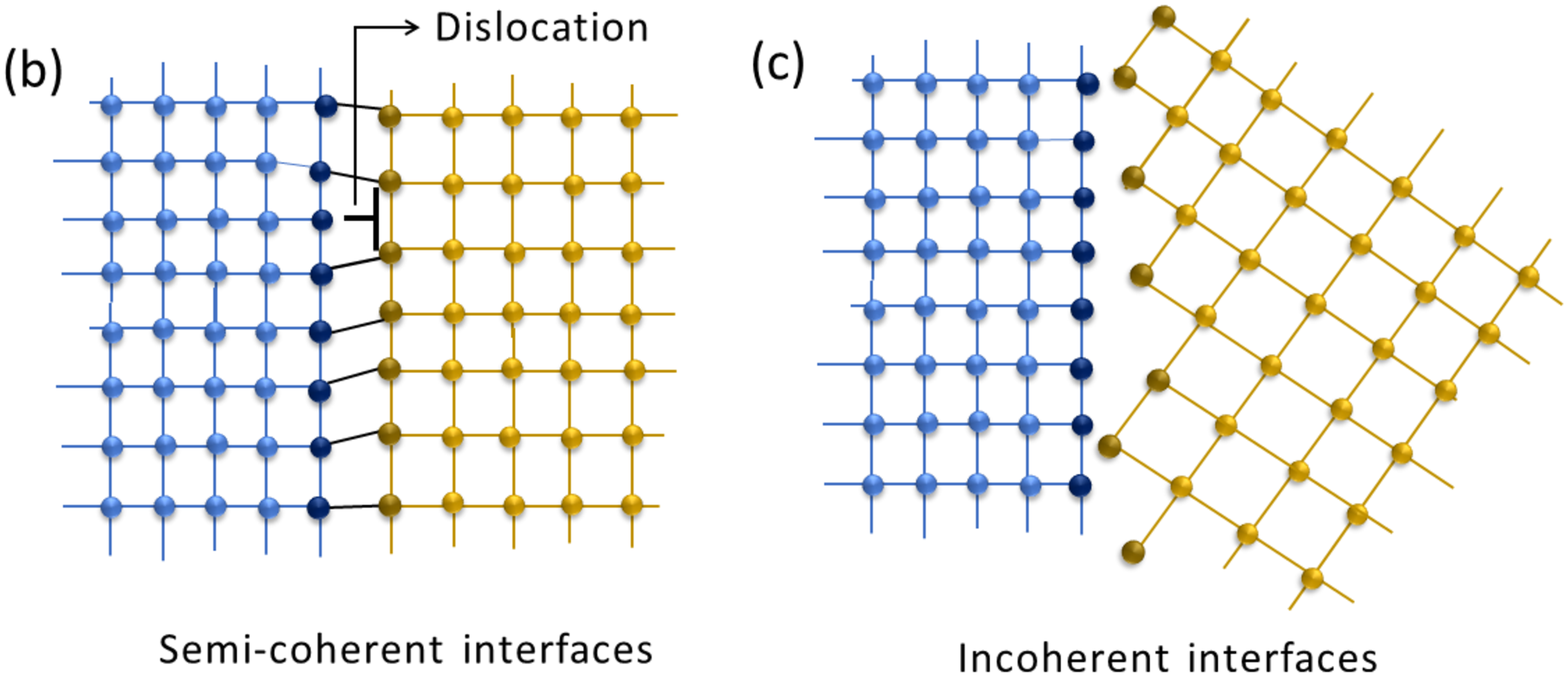}
	\small{\caption{Schematic of (a) polycrysttaline composite containing grains with two different crystalline phases. (b) Dislocations created at the semi-coherent interfaces between two different crystalline materials due to the lattice mismatch, (c) Incoherent interfaces created due to high lattice mismatch between the phases.\label{fig.9}}}
\end{figure} 
When these two different crystal phases are mixed, because of the lattice mismatch they don't fit properly and become distorted as they join at the interface boundaries. This situation helps to create voids and dislocations at the semicoherent or incoherent interfaces, \cite{65} depicted in the schematic of~\Fref{fig.9} (b) and (c). The dislocations are also formed within the planes of crystal structure due to the lattice misfit as pointed in HR-TEM image (\Fref{fig.3}). According to the theoretical prediction \cite{16} and the point contact experiments \cite{66,67}, TLS can be originated from the motion of dislocation segments or the atoms along the grain boundary in polycrystalline materials. Therefore the structural defects created at the interfaces as well as within the crystal planes seem to be responsible for the TLS formation in the present systems. 

\subsection{\label{sec:level12}Coexistence of 2CK with ferrimagnetism}

In magnetic materials the exchange energy splitting can break the symmetry of the two spin channels by changing the electron population density and hence can influence the 2CK behaviour. Therefore the coexistence of ferrimagnetism and orbital 2CK effect in the present composites (x $\geq$ 0.15) leads to an interesting point of discussion. 

Ferrimagnetic materials can be considered as the two interpenetrating sublattices with magnetic moments in the two opposite directions. Due to crystallographic reasons, these two sublattices are not equivalent. Hence their magnetization do not completely cancel each other and we get a net magnetic moment. This explanation suggests that though there will be a net positive magnetization, it will be less than that of a fully polarized state.
\begin{figure}[htb]
	\centering
	\includegraphics[width=1\columnwidth]{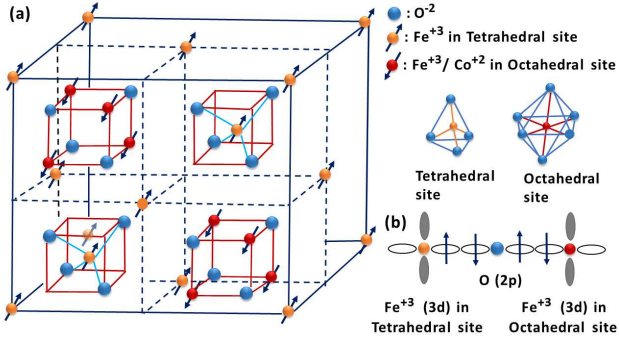}
	\small{\caption{(a) Inverse spinel structure of CoFe$_{2}$O$_{4}$ showing the direction of the spins in octahedral and tetrahedral sites, (b) superexchange coupling between Fe$^{+3}$ ions at octahedral and tetrahedral sites via O$^{-2}$.\label{fig.10}}}
\end{figure}
The net magnetic moment of the magnetic materials depends on the crystallographic structure and cationic distribution. Here, CFO belongs to the spinel structure which consists of two types of lattice sites, tetrahedral and octahedral. In a normal spinel structure (AB$_{2}$O$_{4}$), the  divalent (A$^{+2}$) and trivalent (B$^{+3}$) cations occupy the tetrahedral and octahedral sites respectively. For this type of ideal normal spinel structure the net magnetic moment of CFO is predicted to be 7 $\mu _{B}$ per formula unit (f.u) \cite{68}. But in our system, the value of saturation magnetization of CFO is very small (67.6 emu/g $\approx$ 2.81 $\mu _{B}$/f.u) which means in present case the structure of CFO is not the normal spinel. This small magnetization value corresponds to the inverse spinel where half of the octahedral sites are occupied by Co$^{+2}$ ions and other half as well as all the tetrahedral sites are occupied by Fe$^{+3}$ as shown in~\Fref{fig.10} (a). The Fe$^{+3}$ ions in octahedral sites are aligned antiferromagnetically to that in tetrahedral sites via superexchange interactions mediated by Oxygen (\Fref{fig.10} (b)). Therefore the magnetic moment of Fe$^{+3}$ cations cancel out and the net magnetic moment comes from Co$^{+2}$ ions which gives a value of 3 $\mu _{B}$/f.u due to its three unpaired d-electrons \cite{68,69}. This reduced magnetic moment due to the inverse spinel structure seems to be unable to destroy the channel symmetry and hence could possibly be the reason of unaffected 2CK in our composites. This situation is analogous to the ferromagnetic L1$_{0}$MnGa thin film \cite{11} where the antiparallel alignment between Mn-Mn atoms due to the superexchange coupling reduces the value of M$_{s}$ showing robust existence of the orbital 2CK effect.  

\subsection{\label{sec:level13}Magnetoresistance}

To study the effect of scattering mechanism on magnetotransport properties, we have measured MR at low temperature up to 8 T for the composites with x = 0.10 and 0.15 (\Fref{fig.11} (a) and (b) respectively). The magnetic field is applied perpendicular to the flow of current. Both composites show negative MR and its absolute value decreases consistently with increasing temperature. Although the transport properties are different for these two composites, the MR behaviour is similar. The value of MR is also not affected much with increasing CFO content as clearly depicted from~\Fref{fig.11}.

\begin{figure}[htbp]
	\centering
	\includegraphics[width=1\columnwidth]{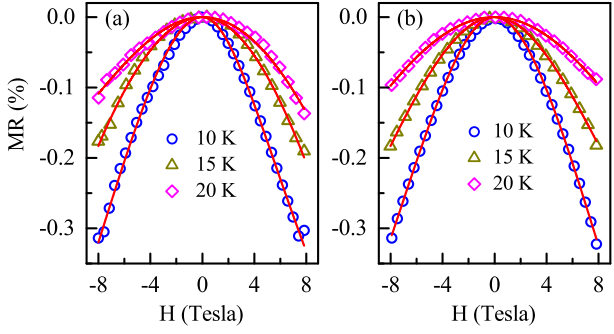}
	\small{\caption{MR at low temperatures for composites with (a) x = 0.10 and (b) x = 0.15\label{fig.11}}}
\end{figure}

In case of ferromagnetic polycrysttaline samples, the spin polarized tunneling at the grain boundaries or the domain wall motion is usually responsible for a negative MR \cite{70,71}. In this case a sharp fall in resistivity is observed at low magnetic field ($<$ 1 T), similar type of behaviour as colossal magnetoresistance.  Since the low field variation as well as the overall value of MR is very less, the grain boundary tunneling is not accountable for negative MR in our samples. Hence we have considered the Khosla Fisher model based on spin dependent scattering to explain the negative MR. According to this model, the localized impurity spins are aligned in the presence of external magnetic field and hence scattering of conduction electrons is reduced resulting a decrease in resistivity \cite{72}. Several groups have successfully explained the negative MR using this model in ferromagnetic materials, such as iron filled multiwalled carbon nanotubes \cite{73}, (Ga$_{1-x}$, Fe$_{x}$)Sb thin films \cite{74} and In$_{1-x}$Mn$_{x}$Sb thin films \cite{75}. The semi-empirical formula of the negative MR as given by Khosla and Fisher can be written as \cite{72}:
\begin{equation}
	\frac{\bigtriangleup\rho}{\rho} = -B_{1}^{2}ln(1+B_{2}^{2}H^{2})
	\label{eq.5}
\end{equation}

This expression is based on third order expansion of the s-d exchange Hamiltonian and the logarithmic temperature variation of resistivity in the dilute magnetic alloy \cite{5,76}.
The coefficients B$_{1}$ and B$_{2}$ are given by:
\begin{equation}
	B_{1}= A_{1}JN(\varepsilon_{F})[S(S+1)+\langle M^{2} \rangle]
	\label{eq.6}
\end{equation}
and
\begin{equation}
	B_{2}^{2}= \Big[1+4S^{2}\pi^{2}\Big(\frac{2JN(\varepsilon_{F})}{g}\Big)^{4}\Big]\Big(\frac{g\mu_{B}}{\alpha k_{B}T}\Big)^{2}
	\label{eq.7}
\end{equation} 

where the parameter $A_{1}$ signifies the contribution of spin scattering to total MR, J is the exchange interaction energy,  N$(\varepsilon_{F})$ is the density of states at Fermi level $(\varepsilon_{F})$, S and g are the total spin aand the effective Land\`{e} factor of localized magnetic moment, $\langle M^{2} \rangle$ is the average of magnetization squared and $\alpha$ is a numerical constant on the order of unity. $A_{1}$ is defined as AN$_{A}$($\sigma_{J}/\sigma_{0}$)$^{2}$ where A is a numerical constant, N$_{A} $ is the Avogadro's number and $\sigma_{J}$ and $\sigma_{0}$ are the scattering cross sections due to exchange interaction and other scattering mechanisms respectively.

The experimental data of our composites is well fitted with~\Eref{eq.5}. It indicates that negative MR is originated due to the reduced interaction of conducting carriers with the localized magnetic moment of CFO. The two fitting parameters B$_{1}$ and B$_{2}$ are listed in~\Tref{table 4} for composites with x = 0.10 and 0.15. The value of B$_{2}$ consistently decreases with increasing temperature due to its 1/T variation whereas change in B$_{1}$ with temperature is non-monotonous. According to~\Eref{eq.7}, B$_{2}$ is proportional to (JN$(\varepsilon_{F})$)$^{2}$. At a fixed temperature, the parameter B$_{2}$ increases with increasing CFO percentage, but the value of B$_{1}$ gets reduced except at 10 K. With increasing CFO content, the product of JN$(\varepsilon_{F})$ increases as evident from the rising value of T$_{K}$ and therefore the parameter B$_{2}$ is also increased. On the other hand, decrease in B$_{1}$ with increasing the amount of CFO could be due to the reduced scattering cross section of exchange interaction and an increase in cross section of other scattering mechanisms such as scattering between the structural defects and the conduction electrons.

\begin{table}[htb]
	\setlength{\tabcolsep}{6pt} 
	\renewcommand{\arraystretch}{1.4} 
	\centering
	\caption{The values of B$_{1}$ and B$_{2}$ obtained after fitting experimental MR data with~\Eref{eq.5} at different temperatures for composites with x = 0.10 and 0.15.\label{table 4}}
	\resizebox{\linewidth}{!}{%
		\begin{tabular}{ccccccc}
			\toprule[0.5pt]
			Sample & \multicolumn{3}{c}{B$_{1}$} & \multicolumn{3}{c}{B$_{2}$ (T$^{-1}$)}\\
			\midrule[0.2pt]
			{}  & 10 K & 15 K & 20 K & 10 K & 15 K & 20 K \\
			\cmidrule[0.2pt](l{0.5em}r{0.5em}){2-4}  \cmidrule[0.2pt](l{0.5em}r{0.5em}){5-7}
			90LN-10CF   &  0.586 & 1.630 & 1.53  & 0.156 &  0.034 & 0.029 \\
			85LN-15CF   &  0.655 & 0.607 & 0.986  & 0.131 &  0.099 & 0.040  \\
			\bottomrule[0.5pt]
		\end{tabular}}
	\end{table}
	
Following above discussion it seems that in our composite system, MR is dominated by the spin disorder scattering of conduction electrons, even in the composite with x = 0.15 instead of orbital 2CK effect as discussed in~\Sref{sec:level10}. There are other examples where the orbital 2CK effect dominates the low temperature transport property, but it does not have any effect on MR, instead MR is influenced by the spin-dependent scattering \cite{26} or by the Lorentz force \cite{18}.

\section{\label{sec:level14}Conclusion}

In summary, we have successfully explained the low temperature transport properties of the composites comprising LaNiO$_{3}$ and CoFe$_{2}$O$_{4}$. The origin of resistivity minimum for lower percentage of CFO (10 \%) is attributed to the spin 1CK effect other than weak localization, EEI or the grain boundary tunneling. On the other hand, for higher content of CFO ($\geq$ 15 $\%$), the magnetic field independent upturn reflects the scattering of conduction electrons with the structural two-level system and hence indicates the orbital 2CK effect. Also the experimental data at constant magnetic field provides evidences for stability of the 2CK parameters such as the tunneling rate of the atom, resonant scattering and the coupling strength between the conduction electrons and TLS against high magnetic field (14 T). The net magnetization of CFO is reduced due to the superexchange coupling between Fe$^{+3}$ ions at the two sublattices in the inverse spinel structure. Therefore the symmetry of the two spin channels is not influenced by the exchange energy splitting. Consequently, the orbital 2CK remains unaffected. A negative MR is manifested at low temperature for both composites with x = 0.10 and 0.15. The negative MR is caused by the localized magnetic moment scattering as interpreted by the Khosla-Fisher model.  

Finally, we emphasize that for lower percentage of CFO (10 \%), the impurity spin controls the electron scattering and hence the conduction mechanism. But when we increase the CFO percentage, it creates more structural defects inside the grains as well as at the grain boundaries. Accordingly, the effect of enhanced interaction between the electrons and the TLS becomes more prominent on the transport properties resulting in orbital 2CK. Thus we have successfully shown a tuning from spin 1CK to orbital 2CK effect in the presence of ferrimagnetism by varying CFO content in the composites with LNO.

\section{\label{sec:level15}Acknowledgements}
	The authors gratefully acknowledge Department of Physics, Centre for Nanoscience and Engineering and Advanced Facilities for Microscopy and Microanalysis at Indian Institute of Science, Bangalore for the support of research facilities and to MHRD, India for research fellowship. The authors are also thankful to Pritam Khan and Narendra Tanty for the helpful discussion.

	\title[]{Tuning spin one channel to exotic orbital two-channel Kondo effect in ferrimagnetic composites of LaNiO$_{3}$ and CoFe$_{2}$O$_{4}$: Supplementary material}
	
	\author{Ananya Patra$^1$, Krishna Prasad Maity$^1$, Ramesh B kamble $^1$$^,$$^2$ and V Prasad$^1$}
	\address{$^1$ Department of Physics, Indian Institute of Science, Bangalore 560012, Karnataka, India}
	\address{$^2$ Department of Physics, College of Engineering, Pune 411005, Maharastra, India}
	\vspace{10pt}
	\begin{indented}
		\item[]April 2018
	\end{indented}
	\submitto{\JPCM}

	\section{\label{sec:level1}Magnetization vs. Temperature}
	\begin{figure}[htb]
		\centering
		\includegraphics[width=0.8\columnwidth]{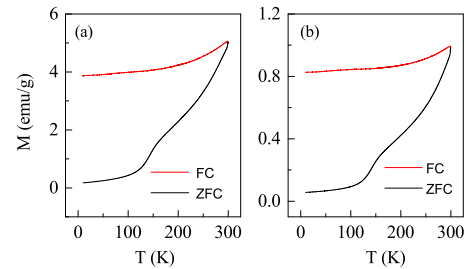}
		\small{\caption{Temperature dependent magnetization in FC and ZFC measurement with H = 50 mT for (a) pure CFO and (b) the composite with x = 0.15 \label{fig.1}}}
	\end{figure} 
	The field cooled (FC) and zero-field cooled (ZFC) magnetization data are recorded with applying magnetic field (H) 50 mT in the temperature range 300--10 K as shown in~\Fref{fig.1}. The temperature variation of magnetization is similar for both the samples, only the magnitude is reduced in the composite. The FC and ZFC curves do not show any bifurcation point in the measured temperature region. The ZFC magnetization curve decreases with decreasing temperature. The FC magnetization curve is almost independent of temperature below 200 K with a initial decrease with lowering temperature. This initial decease could be due to the spin canting where the spins of the metal ions in the two sub-lattices (octahedral and tetrahedral) are not exactly antiparallel to each other. With further decreasing temperature, the thermal fluctuations also get reduced and the spins are aligned in ferrimagnetic manner \cite{1}.

	\section{\label{sec:level2}Calculation of screening constant in three dimension}
	
	Following the calculation of Altshuler and Aronov, the screening factor $\tilde{F}_{\sigma}$ in 3D is given by the expression \cite{2}:
	\begin{equation}
		\tilde{F}_{\sigma} = \frac{32}{3} \frac{[1+\frac{3F}{4}-(1+\frac{F}{2})^{3/2}]}{F}
		\label{eq.1} 
	\end{equation}
	
	where F is related to the screening vector ($\kappa$) and the Fermi wavevector (k$_{F}$) according to Thomas-Fermi approximation \cite{3}:
	
	\begin{equation}
		F = \Big(\frac{\kappa}{2k_{F}}\Big)^{2}ln\Big[1+\Big(\frac{2k_{F}}{\kappa}\Big)^{2}\Big]
		\label{eq.2} 
	\end{equation}

	The ratio of \Big($\frac{\kappa}{k_{F}}$\Big) can be determined from the relation:
	
	\begin{equation}
		\frac{\kappa}{k_{F}} = \Big(\frac{16}{3\pi^{2}}\Big)^{1/3} \Big(\frac{m^{*} r_{s}}{m}\Big)^{1/2}
		\label{eq.3} 
	\end{equation}
	
	The parameter r$_{s}$ is related to the number density of electrons (n):
	
	\begin{equation}
		r_{s} = \Big(\frac{3}{4\pi n}\Big)^{1/3}\frac{1}{a_{0}}
		\label{eq.4} 
	\end{equation}
	
	where a$_{0}$ is Bohr radius.
	
	To get an idea about the value of $\tilde{F}_{\sigma}$, we have used n = 10$^{23}$ cm$^{-3}$ and m$^{*}$ = 11m$_{e}$, the values of pure LNO \cite{4} and we have estimated F = 0.904 which indicates a strong screening according to Thomas-Fermi theory. For F = 0.904,~\Eref{eq.1} gives $\tilde{F}_{\sigma}$ = 0.837.

\end{document}